# Performance Tests of the Kramers Equation and Boson Algorithms for Simulations of QCD

Karl Jansen and Chuan Liu

Deutsches Elektronen-Synchrotron DESY,

Notkestr. 85, D-22603 Hamburg, Germany

Beat Jegerlehner

Max-Planck-Institut für Physik,

Föhringer Ring 6, D-80805 München, Germany

December 12, 1995

## Abstract

We present a performance comparison of the Kramers equation and the boson algorithms for simulations of QCD with two flavors of dynamical Wilson fermions and gauge group $SU(2)$. Results are obtained on $6^3 12$, $8^3 12$ and $16^4$ lattices. In both algorithms a number of optimizations are installed.

# 1 Introduction

One of the most pressing problems in lattice QCD today concerns numerical simulations of dynamical fermions. The CPU-time spent on present computers to generate a statistically independent configuration on reasonably sized lattices (say $16^4$) easily reaches hours. This makes it difficult to obtain sufficient statistics for an accurate determination of relevant physical quantities. It is therefore not surprising that the search for new algorithmic techniques or improvements on existing algorithms is an active research area.

In this letter we compare the performance of the Kramers equation [1, 2] and the boson algorithms [3, 4, 5], both aimed at simulations of dynamical fermions. The performance of an algorithm is the product of the speed of the program, i.e. the CPU-time to generate a configuration (not necessarily an independent one) and the autocorrelation time of a given observable. Of course, the speed of the program depends on the chosen computer architecture. To be complete, one therefore also has to specify the machine on which numerical tests are performed. In our case we used the Alenia Quadrics (APE) massively parallel computers. Although our results will be presented for this particular machine as an example, we will also give a more machine independent measure for the performance in section 3.

Before starting to compare the two algorithms under consideration, we spent some effort to optimize them. These attempts are described for the Kramers equation algorithm in ref.[2] and for the boson algorithm in ref.[5]. We direct the interested reader to these references for further details. Let us here only summarize these works by mentioning that with relatively simple modifications of the algorithms large improvement factors of $O(10)$ can be obtained.

We understand this investigation as only one step of gaining experience with the behavior of the two algorithms, in particular with the relatively new boson algorithm. It is clear that a full QCD simulation is very costly and one should not hope for a similar precision for the autocorrelation time as in, say, spin models. There, algorithms can be tested thoroughly with several million of configurations which even allow for a determination of the critical dynamical exponent [6]. Our aim here is much more moderate. In order to get reliable numbers for the autocorrelation times and estimates of their errors, we stay in a situation where we are not deep in the physically most interesting however numerically very challenging region of chiral symmetry restoration. We hope that enough "statistics" will be gathered by including also results from other studies in the future. For a review of the present status of fermion algorithms see [7].

As in refs. [4, 2, 5], we will study the standard lattice Wilson QCD with gauge group $SU(2)$. We will work on a 4-dimensional euclidean space-time lattice with volume $\Omega = L_s^3 L_t$ and periodic boundary conditions in all four directions. The gauge field $U_\mu(x) \in SU(2)$ lives on the link pointing from $x$ to $x + \mu$, where $\mu = 0, 1, 2, 3$ designates the 4 forward directions in space-time. The quark fields are denoted by $\psi_{Aa\alpha}(x)$ where $A$,$a$ and $\alpha$ are flavor, color and

Dirac indices respectively. The full partition function for the model is given by,

$$\mathcal{Z} = \int \mathcal{D}U\mathcal{D}\bar{\psi}\mathcal{D}\psi \exp\left(-S_g - S_w\right), \tag{1}$$

where the gauge action $S_g$ and the Wilson fermion action $S_w$ are given by:

$$\begin{aligned} S_g &= -\frac{\beta}{2}\sum_P Tr(U_P), \\ S_w &= \sum_x \bar{\psi}(x)(D+m_0)\psi(x). \end{aligned} \tag{2}$$

The Wilson difference operator $D$ which appears in the above expression is given by:

$$\begin{aligned} D &= \frac{1}{2}\sum_\mu \left[\gamma_\mu(\nabla_\mu + \nabla_\mu^*) - \nabla_\mu \nabla_\mu^*\right], \\ \nabla_\mu \psi(x) &= U_\mu(x)\psi(x+\mu) - \psi(x), \\ \nabla_\mu^* \psi(x) &= \psi(x) - U_\mu^\dagger(x-\mu)\psi(x-\mu), \end{aligned} \tag{3}$$

and $U_P$ is the usual plaquette term on the lattice. In the following, we will consider two flavors of Wilson fermions with degenerate masses. Let us write the fermion matrix $M$ in the hopping parameter $\kappa$ representation, $M = 2\kappa(D+m_0)$ with $m_0$ the bare quark mass and $\kappa = (8+2m_0)^{-1}$. In order to go over to the bosonic theory later, we introduce the hermitian matrix

$$Q = \gamma_5 M / \left[c_M(1 + 8\kappa)\right], \tag{4}$$

where $c_M$ is a free parameter of $O(1)$ which will be tuned to optimize the bosonic simulation algorithm. The parameter $c_M$ has to be chosen such that the eigenvalues $\lambda$ of $Q$ satisfy

$$0 < |\lambda| \leq 1. \tag{5}$$

As usual, for the simulations using molecular dynamics algorithms, the fermion determinant is written in terms of Gaussian scalar fields $\phi$ such that the path integral reads

$$\begin{aligned} \mathcal{Z} &= \int \mathcal{D}U\mathcal{D}\phi^\dagger \mathcal{D}\phi\, e^{-S_{eff}}, \\ S_{eff} &= S_g + \phi^\dagger (Q)^{-2}\phi. \end{aligned} \tag{6}$$

## 1.1 Transformation to the bosonic theory

Following ref. [3], the path integral in eq.(6) can be written as a *local* bosonic theory. We start by approximating the function $1/s$ in terms of a series of polynomials $P_n(s)$ of even degree $n$ that satisfy

$$\lim_{n\to\infty} P_n(s) = 1/s \text{ for all } 0 < s \leq 1. \tag{7}$$

In order to apply eq.(7) to $\det Q^2$ as needed for the simulations, we consider the *matrix-valued* polynomials $P_n(Q^2)$. Because of relation (5) the determinant may be written as

$$\det Q^2 = \lim_{n\to\infty} \left[\det P_n(Q^2)\right]^{-1}. \tag{8}$$

The polynomial may be factorized by determining its roots $z_k$, $k = 1, ..., n$

$$P_n(Q^2) \propto \prod_{k=1}^{n} \left[ (Q - \mu_k)^2 + \nu_k^2 \right] , \qquad (9)$$

where

$$\mu_k + i\nu_k = \sqrt{z_k}, \quad \nu_k > 0. \qquad (10)$$

A particular example for such roots will be given below. Using eq.(9), the determinant factorizes accordingly. Integrating the individual determinants in again by means of scalar fields, the partition function becomes

$$\mathcal{Z} = \lim_{n \to \infty} \int \mathcal{D}U \prod_{k=1}^{n} \mathcal{D}\Phi_k \mathcal{D}\Phi_k^\dagger e^{-S_b[U,\Phi]} , \qquad (11)$$

with

$$S_b[U, \Phi] = S_g[U] + \sum_x \sum_k \left[ |(Q - \mu_k)\Phi_k(x)|^2 + \nu_k^2 |\Phi_k(x)|^2 \right] . \qquad (12)$$

Note that this form of the bosonic expression for the QCD path integral leads to a completely local theory in contrast to eq.(6) which is highly non-local due to the appearance of the inverse fermion matrix. The price to pay is obviously the introduction of a number $n$ of scalar field copies. The choice of the polynomial we will be using in this work is as in [4]. It is a Chebyshev polynomial, the roots of which are given by

$$z_k = \frac{1}{2}(1 + \epsilon) - \frac{1}{2}(1 + \epsilon) \cos\left(\frac{2\pi k}{n+1}\right) - i\sqrt{\epsilon} \sin\left(\frac{2\pi k}{n+1}\right) . \qquad (13)$$

It approximates the function $1/s$ uniformly in the interval $\epsilon \leq s \leq 1$. The relative fit error

$$R_n(s) = [P_n(s) - 1/s] s \qquad (14)$$

in this interval is exponentially small:

$$|R_n(s)| \leq 2 \left( \frac{1 - \sqrt{\epsilon}}{1 + \sqrt{\epsilon}} \right)^{n+1} . \qquad (15)$$

We define the accuracy parameter

$$\delta = \max_{\epsilon \leq s \leq 1} |R_n(s)| \qquad (16)$$

which gives a measure of how well the chosen polynomial approximates $1/s$ in the given interval.

## 2 The algorithms

As mentioned in the introduction, we studied two kinds of algorithms, the Kramers equation and the boson algorithm. In the following we discuss the basic ideas of these algorithms and list the improvements implemented.

## 2.1 Kramers equation algorithm

This algorithm has its origin from techniques based on the Langevin equation. It falls into the class of molecular dynamics algorithms and is very closely related to the Hybrid Monte Carlo algorithm [8, 2]. When one considers Brownian motion in an external field, one may describe the evolution of the particle's momentum $p$ and coordinate $q$ separately. The master equation (generalized Fokker-Planck equation) in this situation is called the Kramers equation [9]. The corresponding generalized Langevin form of the Kramers equation may be written as a formal continuum stochastic differential equation

$$\begin{align}
\dot{p} &= -\frac{\delta H}{\delta q} - \gamma p + \eta(t) , \\
\dot{q} &= \frac{\delta H}{\delta p} ,
\end{align} \tag{17}$$

where the stochastic variables $\eta(t)$ are the so-called "white noise" terms. In eq.(17) a fictitious fifth time coordinate $t$ is introduced and $H$ denotes a 4-dimensional Hamiltonian defined by the theory to be considered. The parameter $\gamma$ is a friction coefficient that can be tuned freely.

For numerical simulations the continuous time derivatives are discretized and the time evolution is realized by discrete time integration schemes, using a finite time step (step size) $\epsilon_{md}$, like the standard leapfrog method [8]. The resulting discretized form of eq.(17) finally leads to the Kramers equation algorithm which is an exact algorithm due to the introduction of a global accept/reject step. It was introduced for field theory simulations and tested in simpler models by Horowitz [1], who called it L2MC. In [2] the algorithm was tested the first time for Wilson QCD using $SU(2)$ as the gauge group. It was found that it performs equally well as the Hybrid Monte Carlo algorithm.

The improvements that have been used for this algorithm are even-odd preconditioning [10] and a better leapfrog integration scheme suggested by Sexton and Weingarten [11]. Recently we also started to investigate the biconjugate gradient stabilized algorithm for inverting the fermion matrix [12]. We found that the number of iterations to reach a given residue can decrease by as much as 40% of the number required in the conjugate gradient method. There are alternative possible improvements like the chronological extrapolation method to find better starting vectors for the conjugate gradient algorithm [13]. However, due to the use of large step sizes in the Sexton-Weingarten integration scheme, we do not expect further acceleration of the program in our case. The combination of the polynomial approximation of $1/s$ and the conjugate gradient inversion technique as proposed in [14] seems rather promising but we have not yet tested it.

## 2.2 Boson algorithm

The version of the bosonic algorithm that we use is described in detail in [5]. We took a combination of standard heatbath and over-relaxation techniques to update the gauge and scalar degrees of freedom. In this version several improvements have been implemented. The first is again even-odd preconditioning [10]. The lowest eigenvalue of the preconditioned matrix $\hat{Q}^2$ is larger than the corresponding eigenvalue of $Q^2$. As a result, in a simulation using preconditioning, a smaller number of scalar field copies can be taken. This is very important not only from the point of view of memory requirements. The main improvement comes from the observation [5] that the autocorrelation time depends linearly on the number of scalar fields. Different kinds of mixing of heatbath and over-relaxation updates also lead to substantial improvements on the autocorrelation time. Finally, optimal choices of the parameter $c_M$, see eq.(4), as proposed in [4] and [15] lead to further improvements. Choosing $c_M$ to be less than one raises the lowest eigenvalue of $\hat{Q}^2$ and therefore leads to a smaller number of scalar field copies. On the other hand, one has to make sure that the largest eigenvalue remains below and sufficiently far from one to avoid accidental slow bosonic modes in the simulation.

In practice, the boson algorithm is run with a non-vanishing accuracy parameter $\delta$, eq.(16). However, there have been several proposals to make the algorithm exact [4, 15].

## 2.3 Parameters for the test runs

In this subsection we give explicitly the values of the algorithm parameters that have been used. The performance of the algorithms can react sensitively to changes of these parameters [2, 5]. The tunable parameters in the Kramers equation algorithm are the discrete finite step size $\epsilon_{md}$, the friction coefficient $\gamma$ and a repetition parameter $k$ which determines how often the momenta are refreshed by generating them newly from a Gaussian distribution. We list the actually used parameter values in Table 1 for the three lattice sizes taken, namely $6^3 12$, $8^3 12$ and $16^4$. The parameter $\epsilon_{md}$ is chosen such that the acceptance rate is about 80%.

Table 1: Technical parameters for both algorithms

| Lattice | Kramers | | | Boson | | | |
|---|---|---|---|---|---|---|---|
| | $\epsilon_{md}$ | $k$ | $\gamma$ | $\epsilon$ | $N_{boson}$ | $\delta$ | $c_M$ |
| $6^3 12$ | 0.205 | 3 | 0.5 | 0.01454 | 18 | 2% | 0.6 |
| $8^3 12$ | 0.185 | 4 | 0.5 | 0.0061 | 24 | 4% | 0.745 |
| $16^4$ | 0.125 | 5 | 0.5 | 0.0048 | 44 | 0.38% | 0.7 |

As mentioned in the introduction, the Chebyshev polynomial with roots given in eq.(13) approximates the function $1/s$ in the interval $\epsilon < s \leq 1$ with an exponential fitting error. For the QCD application, this interval is determined by the lowest $\lambda_{\min}$ and largest $\lambda_{\max}$ eigenvalues of the preconditioned matrix $\hat{Q}^2$ [5] that is used in the simulations. It is therefore necessary that before running the boson algorithm an estimate of these two quantities is made.

In Table 2 we list the expectation values of $\lambda_{\min}$ and $\lambda_{\max}$ obtained by the conjugate gradient method [16]. For the boson algorithm, the parameter $\epsilon$ should be chosen to be at the upper edge of the distribution of $\lambda_{\min}$ (see fig.3 in [5]). As mentioned in section 2.2, the parameter $c_M$ was chosen to be less than one. Lowering $c_M$ in Table 1 further for the $8^3 12$ and $16^4$ lattices, while keeping $\epsilon/\langle\lambda_{\min}(\hat{Q}^2)\rangle$ and $\delta$ fixed, does not lead to an improvement in the computational effort. This unexpected phenomenon was first observed in [5] and needs still to be understood. In order to have an agreement with results from HMC runs [4], the accuracy parameter $\delta$ should be at the order of a few percent and the number of scalar fields $N_{boson}$ as given in Table 1 has to be chosen accordingly.

Table 2: Eigenvalue expectation values

| Lattice | $\langle\lambda_{\min}(\hat{Q}^2)\rangle$ | $\langle\lambda_{\max}(\hat{Q}^2)\rangle$ |
| --- | --- | --- |
| $6^3 12$ | 0.0115(4) | 0.9386(4) |
| $8^3 12$ | 0.00539(9) | 0.6100(2) |
| $16^4$ | 0.00478(3) | 0.6961(5) |

## 3 Results

All results that will be presented below are obtained at $\beta = 2.12$ and $\kappa = 0.15$. These parameters correspond to a pion to $\rho$ mass ratio of about 0.95. Simulations at smaller mass ratios are very costly and a reliable determination of the autocorrelation times is difficult [1]. In Table 3 we give the results for several observables as obtained from both algorithms. We measure the plaquette expectation value $\langle P \rangle$, the pion mass $m_\pi$ and the $\rho$-mass $m_\rho$. The masses are determined through

$$\cosh m = \frac{C(L_t/2+1) + C(L_t/2-1)}{2C(L_t/2)} , \qquad (18)$$

with appropriate correlation functions $C(t)$. As was found earlier [4, 17], with the accuracy parameter $\delta$ given in Table 1, the two algorithms give compatible results, although for the $8^3 12$ lattice $\delta$ seems to be too large to reach a real satisfactory agreement for the plaquette expectation value.

We are mostly interested in the question of how much computer time is needed to reach a statistically independent configuration. To answer this question, one needs the speed of the program and most importantly the autocorrelation times. The latter quantity has been obtained by the "window" technique [6]. In this letter we will only compare the integrated autocorrelation time for the plaquette as an example. Other observables, like meson correlation functions or the lowest eigenvalues $\lambda_{\min}$, show similar behavior. The data sample taken was always more than

---
[1] We actually performed runs at a pion to $\rho$-mass ratio of about 0.5 on a $16^4$ lattice. There the autocorrelation time increased substantially and could not be determined reliably. However, the rough estimates that we can obtain in this situation would not lead to a change of the conclusion as given below.

Table 3: Results for both algorithms

| Lattice | Algorithm | $\langle P \rangle$ | $m_\pi$ | $m_\rho$ |
|---|---|---|---|---|
| $6^3 12$ | Kramers | 0.5803(2) | 1.191(8) | 1.275(9) |
|  | Boson | 0.5804(4) | 1.170(12) | 1.254(14) |
| $8^3 12$ | Kramers | 0.5777(3) | 1.052(8) | 1.123(11) |
|  | Boson | 0.5768(2) | 1.044(3) | 1.112(4) |
| $16^4$ | Kramers | 0.5778(1) | 1.003(2) | 1.060(2) |
|  | Boson | 0.5779(1) | 1.004(4) | 1.063(5) |

10 times the measured autocorrelation time. For the Kramers algorithm on the $6^3 12$ lattice and the boson algorithm on the $6^3 12$ and $8^3 12$ lattices, we have run several replica systems which provided completely independent data samples allowing for a reliable error estimation of the autocorrelation times. For the other lattices the total run was blocked into several sub-blocks, each of which again was several times larger than the measured autocorrelation time. Then the results from these sub-measurements were taken for the error analysis. Clearly, in this case the error determination is not as reliable as in the previous case, but it should nevertheless give a reasonable estimate of the error. We performed cross-checks on the integrated autocorrelation times by analyzing the exponential autocorrelation time and inspecting the blocked errors of the observables. We found the results from this analysis to be in agreement with the ones obtained by the window technique.

The smaller $6^3 12$ and $8^3 12$ lattices have been run on the Q1 version of the APE with 8 nodes. The larger $16^4$ lattice has been run on the QH2 version with 256 nodes. In the last two columns of Table 4 we give the autocorrelation time, $\tau[sec]$, in real CPU-seconds taking the speed of the program into account. The subscript $k$ ($b$) stands for the Kramers equation (boson) algorithm. These numbers provide a direct measure of how long to run each algorithm in real time to obtain an independent configuration on which measurements can be performed. We see that for the larger lattices the Kramers equation algorithm appears to be about a factor of 2 better than the boson algorithm. We note that on the $6^3 12$ lattice the situation is reversed which is presumably due to a non optimal tuning of the Kramers equation algorithm parameters.

One may ask the question, whether the results given in Table 4 are specific for the Alenia Quadrics machine. We therefore tried to find a more machine independent criterion. The computationally most expensive part in the Kramers equation algorithm is the conjugate gradient method for the matrix inversion. This inversion, on the other hand, is dominated by matrix $Q$ times vector $\phi$ operations, denoted by $Q\Phi$. Also for the bosonic algorithm similar floating point operations dominate the program [4, 5]. For this reason we give the autocorrelation time $\tau[Q\Phi]$ in units of $Q\Phi$ in columns 3 and 4. We see the same behavior as for real time in terms of this unit, which should give a more machine independent comparison of the performance of the two algorithms.

Table 4: Performance comparison of the Kramers equation and the boson algorithms.

| Lattice | machine size | $\tau_b[Q\Phi]$ | $\tau_k[Q\Phi]$ | $\tau_b[sec]$ | $\tau_k[sec]$ |
|---|---|---|---|---|---|
| $6^3 12$ | Q1 [8 nodes] | 14000(800) | 21000(4000) | 298(20) | 480(100) |
| $8^3 12$ | Q1 [8 nodes] | 36000(2250) | 17000(5000) | 1781(112) | 979(300) |
| $16^4$ | QH2 [256 nodes] | 56000(18600) | 26000(11000) | 990(330) | 540(230) |

# 4 Conclusions

Our conclusions can be summarized by inspecting Tables 3 and 4. The first Table confirms again [4, 17] that we now have two exact methods for simulations of lattice fermions in QCD which give compatible results for important observables of the lattice theory. In particular, the new bosonic algorithm can now provide important cross-checks for results that have been obtained earlier with the Hybrid Monte Carlo method. The bosonic method appears to be practical also on lattices of size $16^4$ where it needs only a moderate number of scalar field copies. It is certainly important that there exist two conceptually very different algorithms leading to the same results in the difficult area of lattice QCD simulations.

Table 4 shows that for the physical situation we have considered here, namely $m_\pi/m_\rho \approx 0.95$, and for the two versions of the algorithms that we have been testing, the boson algorithm is more costly in CPU-time than the Kramers equation algorithm. Taking the errors of the autocorrelation time into account, however, it is also seen that both algorithms perform comparably and are not orders of magnitude different. Given the already long history of the molecular dynamics algorithms, it is more likely that new ways will be found to improve and accelerate the boson algorithm in the future. One attractive possibility is the reject/accept step as proposed first in [15]. In [14] this procedure was already tested (although for a different theory) and encouraging results were reported. It remains to be seen, whether the bosonic algorithm will be advantageous on large lattices due to its better theoretical scaling behavior [3, 7, 17]. Another open question is how severe the problem with lack of reversibility [2, 12] can become for the molecular dynamics algorithms on large lattices.

### Acknowledgements

We thank M. Lüscher for essential comments and very helpful discussions. All numerical simulation results have been obtained on the Alenia Quadrics (APE) computers at DESY-IFH (Zeuthen). We thank the staff of the computer center at Zeuthen for their support.